\documentclass[onecolumn,oneside,a4paper,12pt]{article}

\usepackage[]{exscale}
\usepackage[dvips]{graphicx}
\usepackage{amssymb}

\begin{document}

\title{
Fischler-Susskind holographic cosmology revisited}

\author{
Pablo Diaz\footnote{e-mail: pablo@posta.unizar.es} \quad M. A. Per\footnote{e-mail: maperb@gmail.com}, \quad Antonio Segui\footnote{e-mail: segui@posta.unizar.es} }

\date{
Departamento de Fisica Teorica\\
Universidad de Zaragoza. 50009-Zaragoza. Spain \\[20 pt] }
\maketitle

\vspace{\stretch{1}}

\begin{abstract}
When Fischler and Susskind proposed a holographic prescription based on the Particle Horizon, they found that spatially closed cosmological models do not verify it due to the apparently unavoidable recontraction of the Particle Horizon area. In this article, after a short review of their original work, we expose graphically and analytically that spatially closed cosmological models can avoid this problem if they expand fast enough. It has been also shown that the Holographic Principle is saturated for a codimension one brane dominated Universe. The Fischler-Susskind prescription is used to obtain the maximum number of degrees of freedom per Planck volume at the Planck era compatible with the Holographic Principle.
\end{abstract}

\vspace{\stretch{6}}

\pagebreak

\pagestyle{plain}


\section{Introduction}
One of the most promising ideas that emerged in theoretical physics during the last decade was the Holographic Principle according to which a physical system can be described uniquely by degrees of freedom living on its boundary \cite{tHooft1,Susskinda}. If the Holographic Principle is indeed a primary principle of fundamental physics it should be verified when the entire universe is considered as a physical system. That is, the physical information inside any cosmological domain should be holographically codified on its boundary area. But obviously, if an unlimited region of scale $L$ is considered, its entropy content will scale like volume $L^{3}$ and its boundary area like $L^{2}$; so inevitably the former will grow quicker than the second and the holographic codification will be impossible for big size cosmological domains. The origin of the Holographic Principle is related to black hole horizons; so, it seems natural to relate it now to any kind of cosmological horizon. It is at this stage when the causal relationship that gives rise to cosmological horizons should be taken into account. William Fischler and Leonard Susskind proposed a cosmological holographic prescription based on the particle horizon \cite{FS}
\begin{equation}\label{enun}
  S_{PH}\leq \frac{A_{PH}}{4}.
\end{equation}
The entropy content inside the particle horizon of a cosmological observer cannot be greater than one quarter of the horizon area in Planck units. Enforcing this condition \emph{for the future} of any cosmological model with constant $\omega=p/\rho$  (Friedmann-Robertson-Walker models, FRW) spatially flat, Fischler and Susskind found the limit $\omega <1$. The compatibility of this limit with the dominant energy condition seems to support the Fischler-Susskind (FS) holographic prescription. In section 2, a detailed deduction of this limit is shown. Moreover, the verification of the FS prescription \emph{in the past} is enforced, finding a limit for the entropy density in the Planck era.

\begin{figure}[!hbt]
~\\[0.0cm]
\begin{center}
\includegraphics[width=12cm]{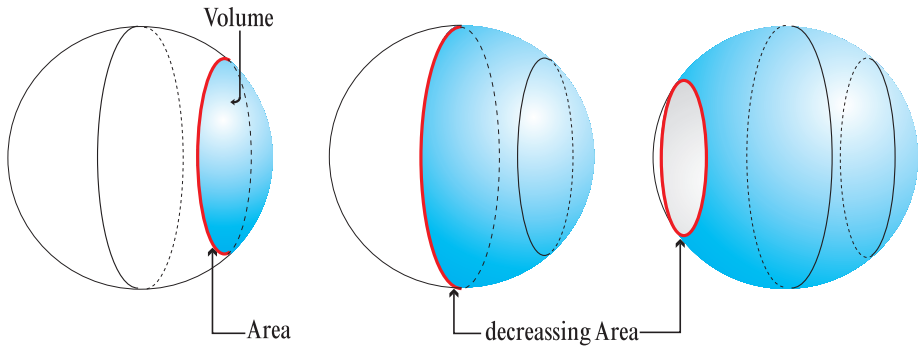}
\end{center}
\caption{{\small Decrease of the area of a domain defined in a compact spatial section when
its volume increases and goes beyond one half of the total volume (further than the
\emph{equator}).}}
~\\[-0.5cm] \label{hpclosed}
\end{figure}

On the other hand, in spatially closed cosmological models, the FS holographic prescription yields to apparently unavoidable problems. Indeed, if the model has compact homogeneous spatial sections, all of them of finite volume, then a physical system cannot have an arbitrary big size at a given time. But for this kind of cosmological models the boundary area does not grow uniformly when the size of a cosmological domain increases. Graphically, it is shown that when the domain crosses the \emph{equator} the boundary area begins to decrease, going to zero when the domain reaches the \emph{antipodes} and covers the entire universe \cite{FS,Bousso2}. Figure~\ref{hpclosed} show this behavior for spatial dimension $n=2$.

Raphael Bousso proposed a different holographic prescription \cite{Bousso2,Bousso1b} based on the evaluation of the entropy content over certain null sections named \emph{light-sheets}. This prescription solves the problems associated to spatially closed cosmological models, but it also lacks the simplicity of the FS prescription. The Bousso prescription will not be used here but it can be shown that both prescriptions are closely related: Two of the \emph{light-sheets} defined by Bousso give rise to the past light cone of a cosmological observer\footnote{According to the Bousso's nomenclature, every past light cone can be built with the light sheets (+-) and (-+) associated to the maximum of that cosmological light cone, also called \emph{apparent horizon} \cite{Bousso2,Bousso1b}.}. According to our previous work \cite{propioBeta}, the entropy content over the past light cone is proportional to the entropy content over the particle horizon (defined over the homogeneous spatial section of the observer), and for adiabatic expansion both will be exactly the same. In fact, the original FS prescription applies to the entropy content over the ingoing past directed null section associated to a given spherical boundary; the key is that the verification for the particle horizon (\ref{enun}) guarantees the verification for every spherical boundary. In conclusion, the FS holographic prescription (\ref{enun}) also imposes a limit on the entropy content over the past light cone, and then it may also be regarded covariant as well as the Bousso prescription.

In section 3 of this paper general explicit solutions for the area and the volume of spherical cosmological domains are obtained in spatially closed (n+1)-dimensional FRW models. It is shown that, in fact, the boundary area of the particle horizon defined in recontracting models (dominated by conventional matter) tends to zero; so, the FS holographic prescription will be violated for this kind of models. But it is also shown that non-recontracting models, that is, spatially closed (n+1)-dimensional FRW models dominated by quintessence matter (bouncing models), do not necessarily present this problematical behavior. These models present accelerated expansion, and particularly only the most accelerated models avoid the collapse of the particle horizon. So, it is deduced that a rapid enough cosmological expansion does not allow the particle horizon to evolve enough over the hyperspheric spatial section to reach the \emph{antipodes}, so the boundary area never decreases. It will be shown that the sufficiently accelerated FRW model corresponds to universes dominated by a codimension one brane gas; thus, such a fluid could saturate the Holographic Principle.

Section 3 concludes with a discussion of our results in contrast with other related works. Especially interesting are the recent works about \emph{holographic dark energy}. The simplified argument is that a holographic limit on the entropy of a cosmological domain could also imply a limit of its energy content; thus, the Holographic Principle applied to cosmology might illuminate the dark energy problem \cite{HDE1,HDE2}. It is argued how our results could improve the compatibility between the particle horizon and the holographic dark energy. Finally, section 4 exposes the basic conclusions of our work.

\section{Fischler-Susskind holography in flat universes}
We will consider (n+1)-dimensional cosmological models with constant parameter $\omega=p/\rho$ (FRW models). Here we study the spatially flat case $k=0$; the scale factor grows according to the potential function
\begin{equation}\label{Rt}
  R(t)\, = \, R_{0} \Big( \frac{t}{t_{0}} \Big) ^{\frac{2}{n(1+\omega)}} \propto t^{1-\frac{1}{\alpha}} \,
\end{equation}
where subscript $0$ refers to the value of a magnitude in an arbitrary reference time $t_{0}$. For later convenience we have defined
\begin{equation}\label{alpha}
  \alpha = \frac{n(1+\omega )}{n(1+\omega )-2}
\end{equation}
$n$ being the spatial dimension of the model. In this section, only conventional matter dominated models --which are decelerated and verify $\alpha >1$-- will be considered, and quintessence dominated models --which are accelerated and verify $\alpha <0$-- are left for the next section. Table~\ref{wacel} summarizes these cases and gives the specific limiting values

\linespread{2.2}
\begin{table}[!h]
~\\[-0.5cm]
\begin{center}
\begin{tabular}{|c|c|c|c|}
\hline
 \quad \textsf{acceleration} \quad &  $\omega$\textsf{-range} & \quad $\alpha$\textsf{-range} \quad & \quad \textsf{denomination} \quad \\
\hline \hline
 $\ddot{R}<0$ & $\displaystyle \frac{2}{n} -1 < \omega \leq +1$ &
 $\displaystyle \ \alpha \geq \frac{n}{n-1} > 0 \ $ & \emph{conventional matter} \\
\hline
$\ddot{R}=0$ & $\displaystyle \ \omega = \frac{2}{n} -1 \ $ & $\alpha =\infty $& \emph{curvature dominated} \\
\hline
 $\ddot{R}>0$ & $\displaystyle \ -1 \leq \omega < \frac{2}{n} -1 \ $ &
 $ \alpha \leq 0$ & \emph{quintessence matter} \\ \hline
\end{tabular}
\end{center}
\linespread{1}\caption{{\small Relation among the cosmological acceleration, the dynamically dominant matter and the parameters of its equation of state $\omega$ and $\alpha$. The ranges can be obtained from the spatially flat case (\ref{Rt}) but they are also valid for the positively (\ref{rclos}) and negatively curved case. The dominant energy condition $| \omega | \leq 1$ and the value $ \omega = -1$ related with a cosmological constant (de Sitter universe) has been also included.}}\label{wacel}
~\\[-0.9cm]
\end{table}
\linespread{1}

Given the scale factor, the particle horizon (named in \cite{HE} like \emph{future event horizon}) for decelerated FRW models can be obtained as \cite{TasiCosmo,EllisHor,Rindler}
\begin{equation}\label{DHP}
  D_{PH}(t)=R(t) \int_{0}^{t}\frac{dt'}{R(t')}=\alpha t \, .
\end{equation}
Assuming adiabatic expansion, the entropy in a comoving volume must be constant; so, the spatial entropy density scales like
\begin{equation}\label{st}
  s(t)R(t) ^{n}=s_{0}R_{0} ^{n}=constant \quad \Rightarrow \quad
  s(t)=s_{0} R_{0}^{n} \, R(t)^{-n}.
\end{equation}
Now the entropy content inside the particle horizon can be computed
\begin{equation}\label{SPH}
  S_{PH}(t) = s(t) V_{PH}(t) =
  s_{0} R_{0}^{n} \, R(t)^{-n} \ \frac{\omega_{n-1}}{n} D_{PH}(t) ^{n} \, ,
\end{equation}
where $\omega_{n-1}$ is the area of the unit sphere. The FS holographic prescription \cite{FS} demands that the above entropy content must not be greater than one quarter of the particle horizon area (\ref{enun}). Then
\begin{equation}\label{APH}
  S_{PH}(t)=s(t) \frac{\omega_{n-1}}{n} D_{PH}(t) ^{n} \leq
  \frac{1}{4}A_{PH}(t) = \frac{1}{4} \omega_{n-1} D_{PH}(t) ^{n-1} \, ,
\end{equation}
performing some cancelations and introducing (\ref{st}) we arrive at
\begin{equation}\label{FS1}
  D_{PH}(t) \leq \frac{n}{4 s(t)} \, = \, \frac{n}{4 s_{0}R_{0} ^{n}}R(t) ^{n} \, .
\end{equation}
This inequality is the simplified form of the FS holographic prescription for spatially flat cosmological models. Now, according to the FS work the inequality should be imposed in the future of any FRW model. For this purpose, comparing the exponents of temporal evolution is sufficient: the particle horizon evolves linearly (\ref{DHP}) and the scale factor evolves according to (\ref{Rt}). Thus, we obtain a family of cosmological models which will verify the FS holographic prescription in the future
\begin{equation}\label{FS2}
  1 \ < \ \frac{2 \ n}{n(1+\omega)}  \quad  \Rightarrow \quad
  \omega < 1 \, .
\end{equation}
This bound on the parameter of the equation of state coincides with the limit of Special Relativity; the sound speed in a fluid given by $v^{2}=\delta p/\delta \rho$ must not be greater than the speed of light. When $\omega=1$, the entropic limit could be also verified depending on the numerical prefactors (see condition (\ref{wlim}) below). So, according to this, the dominant energy condition enables the verification of the FS holographic prescription\footnote{The reverse implication is not valid: the FS prescription allows temporal violations of the dominant energy condition \cite{0306149}.} in the future.

But the previous FS argument presents an objection that we will not obviate. If we enforce that \emph{in the future} the particle horizon area dominates over its entropy content, being potential functions, it is unavoidable that \emph{in the past} the entropy content dominates over the horizon area. In other words, these mathematical functions intersects in a given time, so that at any previous time the holographic codification will be impossible. This intersection time depends on the numeric prefactors that we have previously left out. Our proposal is the enforcement of the intersection time near the Planck time; thus, the apparent violation of the holographic prescription will be restricted to the Planck era. Imposing this limit we will obtain an interesting relation involving the numeric prefactors; so, we have to enforce the simplified holographic relation (\ref{FS1}) at the Planck time ($t_{Pl}=1$). Using (\ref{DHP}) and (\ref{alpha}) we reach
\begin{equation}\label{prefac}
  S_{PH}(t_{Pl})\leq \frac{A_{PH}(t_{Pl})}{4} \quad \Rightarrow  \quad
  \alpha < \frac{n}{4 \, s_{Pl}} \quad \Rightarrow  \quad
  s_{Pl} < \frac{1}{4} \, \big( n-\frac{2}{1+\omega} \big).
\end{equation}
The first idea about this result is that the verification of the Holographic Principle needs, in general, not too high an entropy density; concretely, the FS prescription gives us a limit on the entropy density at the Planck time. This fact is usually skipped in the literature. Perhaps it is assumed that an entropy density at the Planck time $s_{Pl}$ of the same order as one is not problematic. A second view at the previous result may take one to interpret it as a restriction the Holographic Principle imposes on the complexity of our world: the number of degrees of freedom per Planck volume at the Planck era must not be greater than the previous value. Thus, taking $n=3$ and assuming a radiation dominated universe ($\omega=1/3$) at early times, we get $s_{Pl} < 3/8$. Note also that this result does not depend on the final behavior of the model, in a way that is also valid for our universe which is supposed to be dominated now by some kind of dark energy.

Restriction (\ref{prefac}) is not trivial. If we consider a cosmological model dynamically dominated by a fluid with $\omega$ very near to the limit
\begin{equation}\label{wlim}
  \omega_{lim}=\frac{2}{n}-1 \quad \quad ( \, \alpha=\infty \, ) \, ,
\end{equation}
then, the entropy density required at Planck time (\ref{prefac}) will be absurdly small. This is because the models with fluid of matter driven by (\ref{wlim}) do not present particle horizon ($R(t) \propto t$); near this limit the particle horizon becomes arbitrarily big, so the entropy content --scaled with the volume-- can hardly be codified on the horizon area. Moreover, according to \cite{lin} the observational data are compatible with a universe very near the linear evolution; so this case cannot be discarded.

Bousso \cite{Bousso2}, Kaloper and Linde \cite{KL} proposed an \emph{ad hoc} solution based on a redefinition of the particle horizon. They took integral (\ref{DHP}) from the Planck time $t=1$ instead of $t=0$ as the starting point. However, it is not a valid solution for accelerated models ($\omega < \omega_{lim} \ \sim \ \alpha <0$); let us see the reason. According to the new prescription, the redefined particle horizon $\tilde{D}_{PH}$ grows as the scale factor (\ref{Rt})
\begin{equation}\label{DHP2}
  \tilde{D}_{PH}(t)=R(t) \int_{1}^{t}\frac{dt'}{R(t')}=\alpha ( t - t^{1-1/\alpha}) \sim
  -\alpha \ t^{1-1/\alpha} \, .
\end{equation}
So, computing the associated entropy content $\tilde{S}_{PH}$ --with the entropy density (\ref{st})-- leads to a function that approaches a constant value; it can be simplified taking the Planck time as reference time
\begin{equation}\label{SPH2}
  \tilde{S}_{PH}(t) =
  s_{0} R_{0}^{n} \, R(t)^{-n} \, \frac{\omega_{n-1}}{n} \tilde{D}_{PH}(t) ^{n}
  \quad  \Rightarrow \quad \lim_{t \rightarrow \infty} \tilde{S}_{PH}(t) =
   \frac{\omega_{n-1}}{n} s_{Pl} | \alpha | ^{n} \, .
\end{equation}
This limit for the entropy content seems fairly unnatural because it is of the same order as one.

\section{Fischler-Susskind holography in closed universes}

Let us focus on Robertson-Walker metrics with closed spatial sections (curvature parameter $k=+1$). The line element in conformal coordinates ($\eta, \chi$) reads
\begin{equation}\label{rw}
  ds^{2}=R^{2}(\eta) \big(  -d\eta^{2}+ d \chi^{2}+\sin^{2}(\chi)d\Omega^{2}_{n-1} \big) \, ,
\end{equation}
where $d\Omega_{n-1}$ is the metric of the (n-1)-dimensional unit sphere. The inner volume and area of a spherical domain of coordinate radius $\chi$ can be obtained by integrating this metric at a given cosmological time
\begin{equation}\label{arean}
  A(\eta,\chi)= \omega_{n-1} \, R(\eta)^{n-1} \sin^{n-1}(\chi)
\end{equation}
\begin{equation}\label{vol}
  V(\eta , \chi)=R(\eta)^{n} \omega_{n-1} \int_{0}^{\chi} \sin^{n-1}(\chi ') \, d \chi' \, .
\end{equation}
The entropy content inside this volume is obtained using the entropy density (\ref{st})
\begin{equation}\label{Sclosed}
  S( \chi)=s_{0}R_{0} ^{n} \, \omega_{n-1} \int_{0}^{\chi} \sin^{n-1}(\chi ') \, d \chi' \, ,
\end{equation}
where scale factors $R(t)$ have been cancelled; thus, the entropy content inside a comoving volume is constant (adiabatic expansion). Note that $S( \chi)$ strictly grows with the conformal size $\chi$ of the spherical domain; however boundary area $A(\eta,\chi)$ reaches a maximum near \emph{the equator}: for $\chi > \pi /2 $ the boundary area decreases, going to zero at the \emph{antipodes}, where $\chi \rightarrow \pi$ (see Fig.~\ref{hpclosed}). Similar problems appear when the cosmological model recontracts to a Big Crunch, because every boundary area will shrink to zero. In both cases holographic codification will be impossible. This problem will be reviewed in detail and a solution based on the cosmological acceleration will be proposed in the next section.

\subsection{Conventional matter dominated cosmological models}

Fischler and Susskind applied the previous ideas to a FRW (3+1)-dimensional spatially closed cosmological model, dynamically dominated by conventional matter \cite{FS}; the explicit solution for the scale factor is
\begin{equation}\label{rclos}
  R(\eta)=R_{m}\Big| \sin \frac{\eta}{\alpha -1} \Big| ^{\alpha -1} \, .
\end{equation}
Here $R_{m}$ is the maximum value of the scale factor on decelerated models ($\alpha>1$ for conventional matter, see Table~\ref{wacel}); it depends on the relation $\Omega$ between the energy density of the model and the critical density
\begin{equation}\label{Rm}
  R_{m} \equiv R_{0}  \Big( \frac{k}{1-\Omega_{0}^{-1}} \Big) ^{\frac{\alpha -1}{2}}.
\end{equation}
Introducing this scale factor on (\ref{arean}), and computing (\ref{Sclosed}) for the usual case $n=3$, the relation between the entropy content and the boundary area of a spherical domain of coordinate size $\chi$ at the conformal time $\eta$ is obtained
\begin{equation}\label{safs}
  \frac{S}{A}(\eta, \chi )=\frac{s_{0}R_{0}^{2}}{2 R^{2}_{m}} \ \frac{2 \chi - \sin 2 \chi}
  {  ( \sin\frac{\eta}{\alpha -1} )  ^{2(\alpha -1)} \sin ^{2} \chi }.
\end{equation}
It should also be kept in mind that the maximum domain accessible at a given time $\eta$ is the particle horizon; so this relation must be evaluated for $\chi_{PH}(\eta)$, the value that locates the particle horizon for each $\eta$ \cite{TasiCosmo,Rindler}
\begin{equation}\label{chiHP}
  \chi _{PH}(\eta)=\eta -\eta_{BB},
\end{equation}
where $\eta_{BB}$ is the value of the conformal time assigned to the beginning of the universe (usually the Big Bang). A quick observation of relation (\ref{safs}) shows that the denominator goes to zero at $\chi _{PH}= \pi$ (\emph{antipodes}) and also when the scale factor collapses in a Big Crunch; for both cases the ratio $S_{PH}/A_{PH}$ diverges and so the holographic codification (\ref{enun}) is impossible. All FRW spatially closed dynamically dominated by conventional matter models (that is $-1/3<\omega\leq 1$ for $n=3$) will finally recollapse; so, these models will violate the FS holographic prescription.

\subsection{Quintessence dominated cosmological models}

As seen in the last section, some scenarios can become problematic for the holographic prescription. This section aims to expose an alternative solution for some of those troubling cosmological models. The key point in what follows lies in the fact that not all spatially closed cosmological models do recollapse; for example a positive cosmological constant could avoid the recontraction and finally provide an accelerated expansion. The same can be said for different mechanisms which drive acceleration. The present study provides an example where the final accelerated expansion is driven by a negative pressure fluid; this means considering FRW spatially closed (curvature parameter $k=+1$) cosmological models dynamically dominated by quintessence matter, that is $\alpha <0$ (see Table~\ref{wacel}).

The explicit solution for this kind of models is (\ref{rclos}) as well, but its behavior is very different: a negative exponent for the scale factor prevents it from reaching the problematic zero value and so these models are safe from recollapsing in a Big-Crunch and from presenting a singular Big-Bang. Now, the scale factor take a minimum value at same $\eta$; firstly the universe contracts, but after this minimum it undergoes an accelerated expansion for ever; these are called \emph{bouncing models} \cite{Bounc4}. Bouncing models present the obvious advantage of being free of singularities \cite{BekBounc}, and they also enjoy a renewed interest \cite{Bounc3} due to the observed cosmological acceleration \cite{0409088} and especially in relation with brane-cosmology \cite{Bounc4}\footnote{However, our simplest bouncing models associated to the general solution (\ref{rclos}) usually are not considered in the literature.}. On the other hand bouncing cosmologies meets with many problems when trying to reproduce the universe we observe; so the solution (\ref{rclos}) must be only considered like a toy model to study the final behavior of an spatially closed and finally accelerated cosmological model.  Now, formula (\ref{Rm}) gives the minimum value of scale factor $R_{m}$, and according to it $R_{m}$ tends to zero when the energy density tends to the critical density ($\Omega \rightarrow 1$). For an almost flat bouncing cosmology, near the minimum on the scale factor $R_{m}$ quantum gravity effects could dominate erasing every correlation coming from the previous era\footnote{George Gamow words refering to bouncing models: ``from the physical point of view we must forget entirely about the precollapse period'' \cite{gamow}.}. So, in following calculations the beginning of the cosmological time is going to be taken at the minimum on the scale factor (like a no-singular Big-Bang); according to (\ref{rclos}), this corresponds to a conformal time $\eta _{BB}=\pi (1- \alpha ) /2 $. The coordinate distance to the particle horizon (\ref{chiHP}) is then
\begin{equation}\label{chifs}
  \chi _{PH}(\eta )=\eta -\eta_{BB}=\eta -\frac{\pi}{2} (1- \alpha ) \, .
\end{equation}
It was also obtained from (\ref{rclos}) that the scale factor diverges for $\eta _{\infty}=\pi( 1- \alpha )$. This bounded value of the conformal time implies a bounded value for the coordinate size of the particle horizon $\chi _{PH}(\eta_{\infty})$ too.

\begin{figure}[!p]
~ \\[0.0cm]
\begin{center}
\includegraphics[width=14cm]{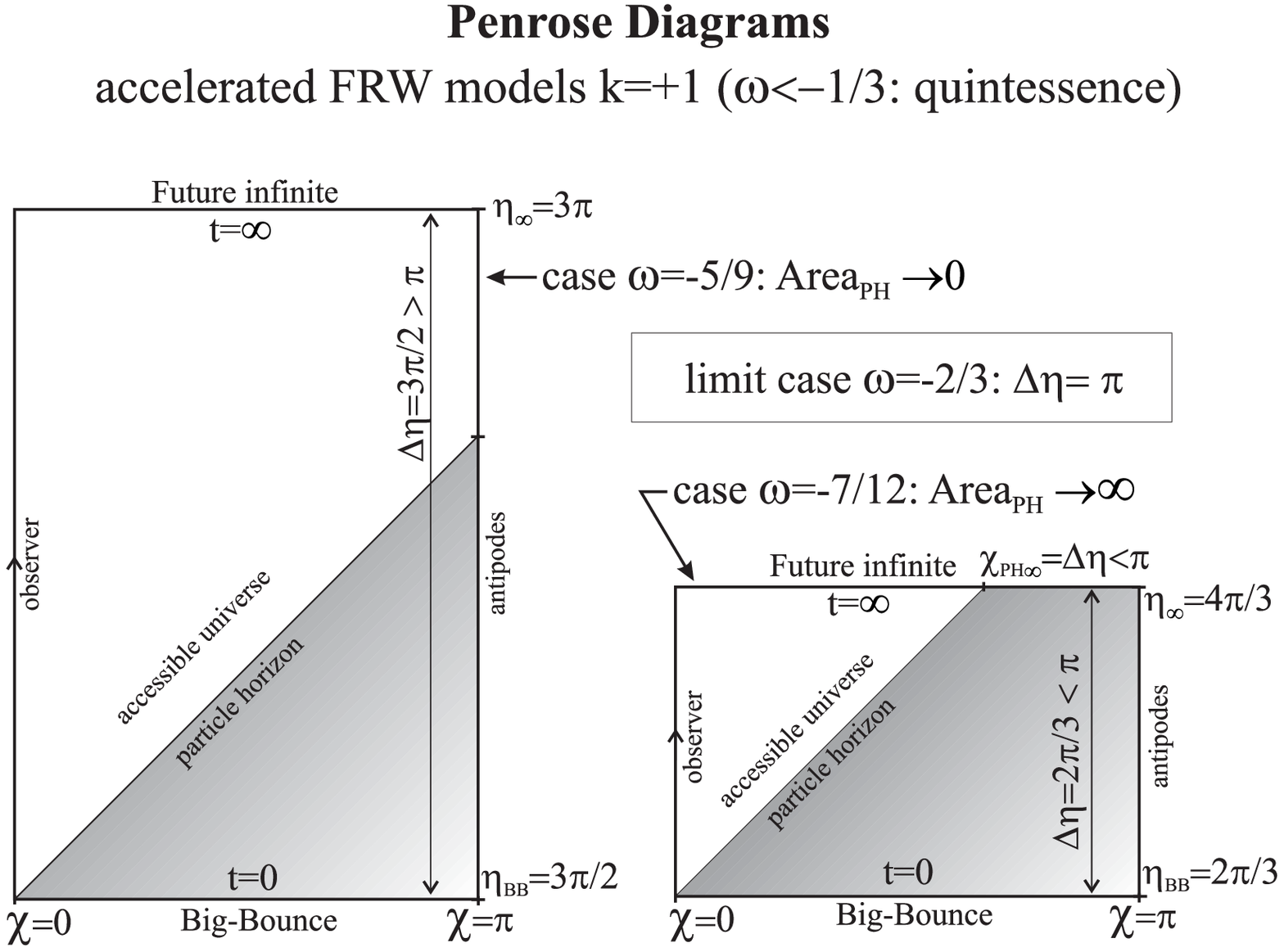}
\end{center}
~ \\[-0.5cm]
\caption[]{{\small Penrose diagrams for spatially closed FRW universes dominated by quintessence (spatial dimension $n=3$); at the ``Big-Bounce'' the scale factor reaches a minimum but at the ``future infinite'' diverges. Depending on the particle horizon behavior two very different cases are shown:\\
$\bullet$ On the left the particle horizon reaches the antipodes $\chi=\pi$; in this case the particle horizon area firstly grows but later it surpasses the equator of the hyperspherical spatial section and finally decreases and shrinks to zero (see Fig.~\ref{hpclosed}) in a finite time. In this case the holographic codification will be impossible.\\
$\bullet$ But on the right the model is more accelerated and so the scale factor diverges for a lower value of the conformal time; so the diagram height is shorter and the particle horizon cannot reach the antipodes. In this case the particle horizon area diverges (due to the divergence of the scale factor at the future infinite) and the holographic codification is always possible.\\
The height of diagram $\Delta \eta$ discriminates both behaviors; so, the limit case is obviously $\Delta \eta = \pi$; then the limit value $\omega = -2/3$ is obtained. For this limiting case the particle horizon reaches the antipodes at the future infinite; the scale factor diverges, the particle horizon area also diverges and, as a consequence, the holographic codification is allowed. So, the $\omega$-range compatible to the holographic codification on the particle horizon is $-1\leq \omega \leq -2/3$ which corresponds to very accelerated spatially closed cosmological models. In general, a sufficient cosmological acceleration do not permit the recontraction of the particle horizon at the antipodes and enables the Fischler-Susskind holographic prescription.}}
~\\[0.9cm]
\label{HorPen}
\end{figure}

\begin{figure}[!p]
~\\[-0.0cm]
\begin{center}
\includegraphics[width=11cm]{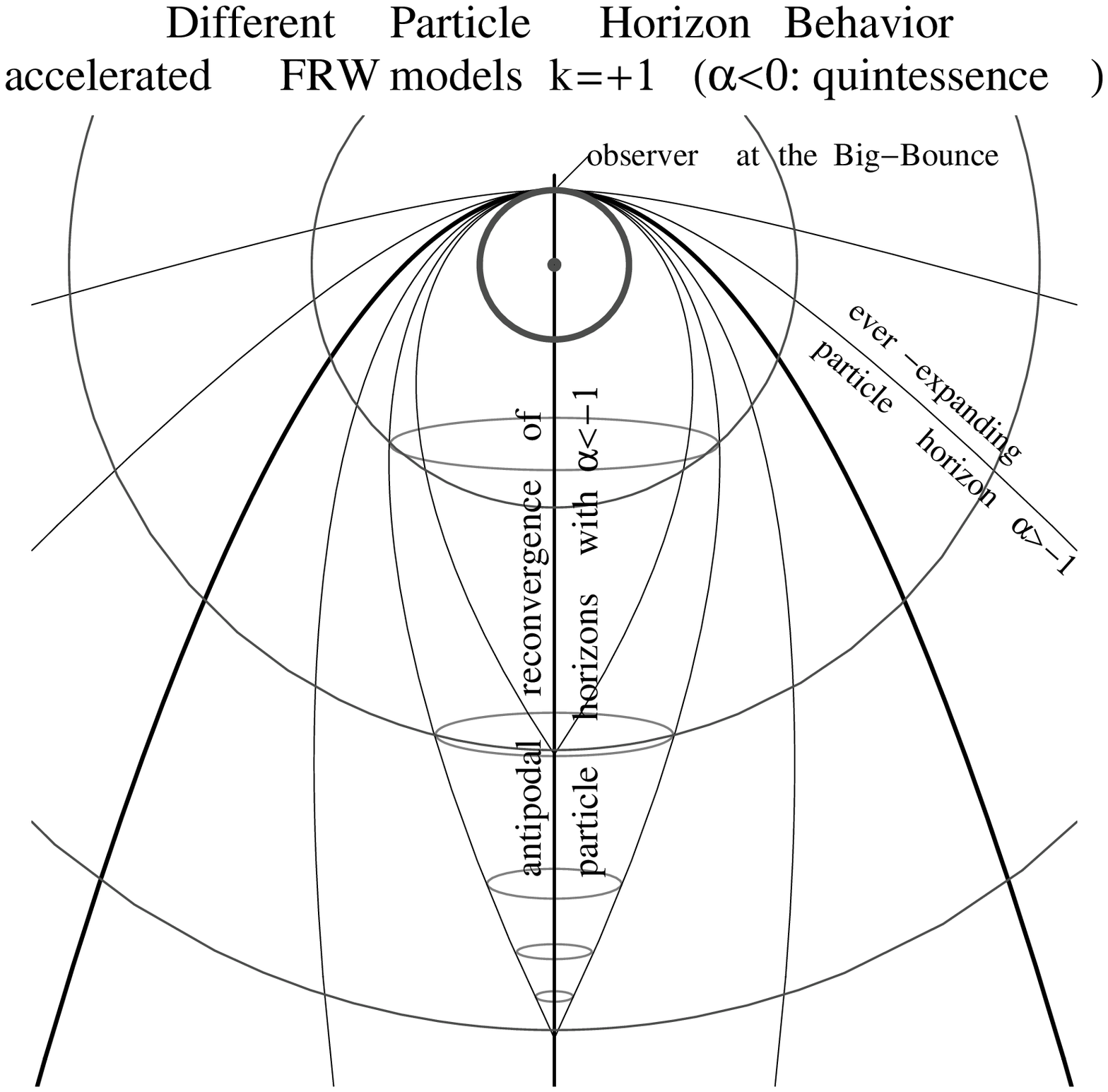}
\end{center}
\caption[]{{\small Polar representation of particle horizons for quintessence dominated ($\alpha <0$) spatially closed FRW models. Future light cones are represented from the beginning $\eta=\eta_{BB}$ (Big-Bounce) for an observer at $\chi=0$. For $\alpha <-1$ the particle horizon reconverges in the antipodes (it reaches and surpasses value $\chi=\pi$), so the particle horizon area shrinks to zero; this shrinkage for a particular future light cone is also shown in the figure. However, for $\alpha \geq-1$ the particle horizon does not reconverge since the cosmological acceleration does not allow it. The FS holographic prescription would be verified in this case. A thick line has been used to show the limit case $\alpha =-1$ ($\omega =-2/3$ if $n=3$).\\
The accelerated growth of the closed spatial sections (3-spheres) is shown by concentric circles; the smallest of them is considered the beginning of the universe, so all the particle horizons (future light cones) arise from it. In this kind of representations the radial distance coincides with the physical radius of the spatially closed model. So, in the figure, light cones do not show the usual 45 degrees evolution. In fact, at the beginning, the future light cones are very flattened since the scale factor of bouncing models evolves very slowly near the minimum which is considered the beginning of time.}} \label{zefscer}
~\\[-1.0cm]
\end{figure}

As argued before, problems for the FS holographic prescription arise at $\chi _{PH}=\pi$, i. e. the value at which a refocusing of the particle horizon on the antipodes of the observer takes place (the horizon area goes to zero). However, this scenario can be avoided by preventing the conformal time from reaching the problematic value (see Fig.~\ref{HorPen}); such FRW spatially closed models will never present any particle horizon recontraction
\begin{equation}
  \chi _{PH\infty}< \pi \quad \Leftrightarrow \quad
  \eta _{\infty}-\eta_{BB}=\frac{\pi}{2} (1- \alpha )< \pi
  \quad \Leftrightarrow \quad \alpha >-1 \, .
\end{equation}
Quintessence models also verify $\alpha <0$; then the allowed range becomes $0>\alpha >-1$ which corresponds to very accelerated cosmological models.

This result can be physically interpreted as follows: For very accelerated spatially closed cosmological models the growing rate of the scale factor is so high that it does not permit null geodesics to develop even \emph{half a rotation} over the spatial sections (see Fig.~\ref{zefscer}). So the particle horizon, far from reaching the antipodal point, presents an eternally increasing area. It also happens for the limiting case $\alpha =-1$ ($\omega = -2/3$ if $n=3$) due to the divergence of the scale factor. This can be summarized in the next statement: \emph{every spatially closed quintessence model with $\alpha \geq -1$ has an eternally increasing particle horizon area}.

\begin{figure}[!tb]
~ \\[0.0cm]
\begin{center}
\includegraphics[width=15cm]{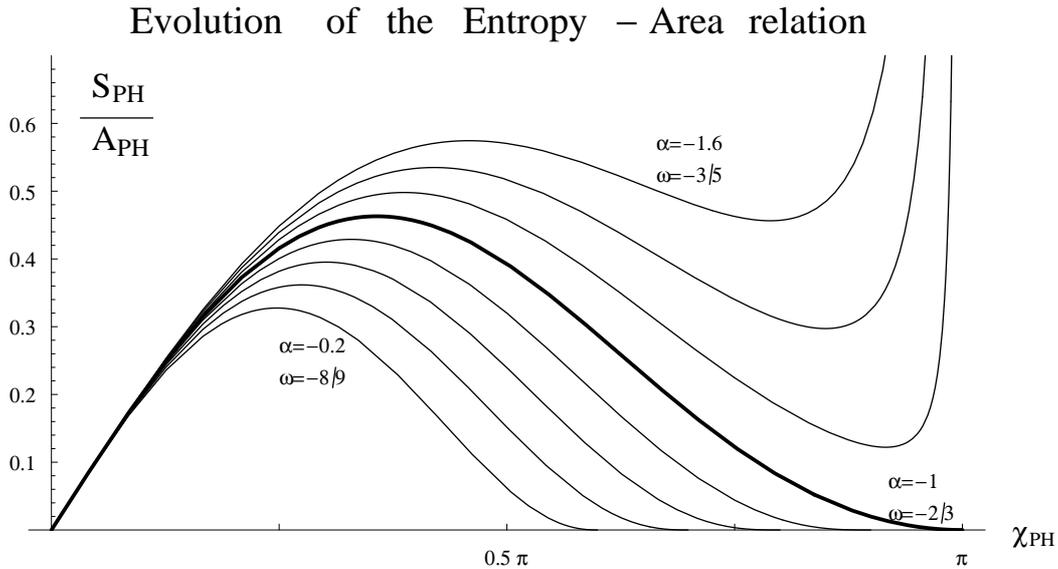}
\end{center}
\caption{{\small Evolution of quotient $S_{PH}/A_{PH}$ depending on the coordinate distance $\chi_{PH}$ as the particle horizon evolves and assuming $s_{m}=1$. Functions for different values of the parameter  $\alpha(\omega)$ are shown. A thick line represents the limit case $\alpha =-1$. For $\alpha <-1$ ($\omega >-2/3$ if $n=3$) the quotient diverges as the particle horizon reaches $\chi_{PH}=\pi$ (the particle horizon area shrinks to zero at the \emph{antipodes} of a fiducial observer). But for very accelerated models, $\alpha \geq-1$ ($\omega \leq-2/3$ if $n=3$), the quotient is always finite which is a necessary condition for the FS holographic prescription to be verified.}}
~\\[0.9cm]
\label{zefs}
\end{figure}

The volume of the spatial sections for spatially closed cosmological models is always finite, and so the entropy content will be; moreover the entropy content of the universe for adiabatic expansion is constant. Then, in accordance with the previous result, the relation $S_{PH}/A_{PH}$ remains finite and goes to zero (see Fig.~\ref{zefs}); now, using (\ref{alpha}) leads to the conclusion that the FS holographic limit is also compatible with FRW spatially closed models verifying
\begin{equation}\label{wcri}
  \omega \, \leq \, \frac{1}{n}-1 \quad \quad (n=3, \ \ \omega \leq -\frac{2}{3} \ ) .
\end{equation}
D. Youm \cite{Youm} applies the same argument to brane universes and arrives to similar conclusions. Note that the limiting value $\omega = \frac{1}{n}-1$ corresponds to a gas of co-dimension one branes \cite{0506053}; with this kind of matter the FS holographic limit could be saturated depending on the numerical prefactors (like the value of the entropy density $s_{0}$).

The FS prescription is neither violated in the past since entropy content $S_{PH}$ goes to zero quicker than the particle horizon area $A_{PH}$ as the beginning is approached, in a way that the relation $S_{PH}/A_{PH}$ also goes to zero. This behavior may be checked by introducing (\ref{chifs}) in the general equation (\ref{safs})
\begin{eqnarray}
  \frac{S_{PH}}{A_{PH}}(\chi_{PH} )& = & s_{m} \ \frac{\chi_{PH}-\sin \chi_{PH} \cos \chi_{PH}}{\sin ^{2} \chi_{PH}}
  \Big(  \cos \frac{\chi_{PH}}{1-\alpha }  \Big) ^{2(1- \alpha ) } \label{safsq2} \\
  \chi_{PH} \ll \pi \, : \quad \quad & \simeq & \frac{2}{3}s_{m} \, \chi_{PH} \, ,
\end{eqnarray}
where $s_{m}$ is the spatial entropy density at the beginning of the universe, which is chosen as reference time (so $s_{0}=s_{m}$ and $R_{0}=R_{m}$). Fig.~\ref{zefs} shows function (\ref{safsq2}) for different values of $\alpha(\omega)$; there, the behavior that has been analytically deduced may be graphically verified. Looking at maxima of the $S_{PH}/A_{PH}$ functions proves that, for non-problematic cases ($\alpha \geq -1$), value $0.5$ is an upper bound, so that
\begin{equation}
  \alpha \geq -1 \ \ ( n=3, \ \omega \leq -2/3) \quad \Rightarrow \quad
  \frac{S_{PH}}{A_{PH}}(\eta )< 0.5 \, s_{m} \, .
\end{equation}
The maximum initial entropy density compatible with the FS entropic limit depends on this bound and this turns out to be
\begin{equation}\label{prefac2}
  s_{m} \leq 1/2 \quad  \Rightarrow \quad S_{PH} \leq \frac{A_{PH}}{4} \, .
\end{equation}
This means that to impose not to have more than one degree of freedom for each two Planck volumes is enough to ensure the verification of the FS prescription for spatially closed and accelerated FRW models with $\alpha > -1$.

\subsection{A more realistic cosmological model}

The previous results are based on a simple explicit solution for the scale factor (\ref{rclos}) but its beginning (the bounce) probably is far from the real evolution of our universe. Here the opposite point of view is exposed: a two-fluid explicit, but not simple, solution mimics a spatially closed cosmological model according to the observed behavior. The Friedmann equations with curvature parameter $k=+1$ can be solved exactly for a universe initially dominated by radiation plus a positive cosmological constant $\Lambda$ that finally provides the desired final acceleration\footnote{For a small enough $\Lambda$ the attractive character of the radiation always dominates and the universe recollapses in a Big-Crunch. Like in the classical Lema\^itre's model (initially dominated by pressureless matter) there exists a critical value $\Lambda_{c}$ which provides a static but inestable model.}. The scale factor then evolves as
\begin{equation}\label{radcc}
  R(t) \ = \ \frac{1}{2\sqrt{\frac{\Lambda}{3}}} \ \sqrt{2-2 \cosh \big( 2 \sqrt{\frac{\Lambda}{3}} \, t \big) +
  4\sqrt{C_{\gamma} \frac{\Lambda}{3}} \, \sinh \big(2 \sqrt{\frac{\Lambda}{3}} \, t \big)} \, ,
\end{equation}
where $C_{\gamma}$ is a constant related to the radiation density $\rho_{\gamma \, 0}$ measured in an arbitrary reference time: \begin{equation}\label{AA}
  C_{\gamma}  \ = \ \frac{8 \pi}{3} \ \rho_{\gamma \, 0}\, R_{0}^{4} \, .
\end{equation}
Due to the initial deceleration (radiation dominated era) this model presents a genuine particle horizon defined by the future light-cone from the Big-Bang. The evolution of this light-front over the compact spatial sections is better described by the conformal angle
\begin{equation}\label{chiPH}
  \chi_{PH}(t) \ = \ \int^{t}_{0} \, \frac{dt}{R(t)} \ .
\end{equation}
Like in the previous section if this conformal angle reaches the value $\pi$ for a finite time this means that the particle horizon has covered all the spatial section, that is, it has reached the antipodes. There the particle horizon area is zero and the FS holographic prescription is not verified. But the proposed model is finally dominated by a positive $\Lambda$ that provides an extreme (exponential) cosmological acceleration that could prevent the refocusing of the particle horizon. It can be checked that the conformal angle never reaches the problematic value $\pi$ when the parameters verify $C_{\gamma} \Lambda > 1.2482 $ (in Planck units).

Experimental measurements suggest that our universe is flat or almost flat; here the second case is assumed, based on the value $\Omega=1.02\pm0.02$ from the combination of SDSS and WMAP data \cite{Uflat5}. The best fit of the scale factor (\ref{radcc}) to the standard cosmological parameters $H_{0}$, $t_{0}$ and $\Omega_{\Lambda}$ takes place for $C_{\gamma} \Lambda \sim 700 $. Thus, the final acceleration of our universe seems to be enough to avoid the refocusing of the particle horizon; particularly it will tend to the asymptotic value $\chi_{PH\infty}\sim 0.5 \, rad$. The conclusion is that if our universe is positively curved and its evolution is similar to (\ref{radcc}) then it could verify the FS holographic prescription far from saturation due to the ever increasing character of the particle horizon area.

\subsection{Discussion and related works}

After the Fischler and Susskind exposition of the problematic application of the holographic principle for spatially closed models \cite{FS} and R. Easther and D. Lowe confirmed these difficulties \cite{EaLowe}, several authors proposed feasible solutions. Kalyana Rama \cite{KR} proposed a two-fluid cosmological model, and found that when one was of quintessence type, the FS prescription would be verified under some additional conditions. N. Cruz and S. Lepe \cite{CruzLepe} studied cosmological models with spatial dimension $n=2$, and found also that models with negative pressure could verify the FS prescription. There are some alternative ways such as \cite{0306149} which are worth quoting. All these authors analyzed mathematically the functional behavior of relation $S/A$; our work however claims to endorse the mathematical work with a simple picture: ever expanding spatially closed cosmological models could verify the FS holographic prescription, since, due to the cosmological acceleration, future light cones could not reconverge into focal points and, so, the particle horizon area would never shrink to zero.

As one can imagine, by virtue of the previous argument there are many spatially closed cosmological models which fulfill the FS holographic prescription; ensuring a sufficiently accelerated final era is enough. Examples other than quintessence concern spatially closed models with conventional matter and a positive cosmological constant, the so-called \emph{oscillating models of the second kind} \cite{Narlikar}. In fact, the late evolution of this family of models is dominated by the cosmological constant which is compatible with $\omega =-1$, and this value verifies (\ref{wcri}). Roughly speaking, an asymptotically exponential expansion will provide acceleration enough to avoid the reconvergence of future light cones.

One more remark about observational result comes to support the study of quintessence models. If the fundamental character of the Holographic Principle as a primary principle guiding the behavior of our universe is assumed, it looks reasonable to suppose the saturation of the holographic limit. This is one of the arguments used by T. Banks and W. Fischler \cite{banks2,banks3} to propose a holographic cosmology based on a an early universe, spatially flat, dominated by a fluid with $\omega =1$\footnote{Banks and Fischler propose a scenario where black holes of the maximum possible size --the size of the particle horizon-- coalesce saturating the holographic limit; this ``fluid'' evolves according to $\omega =1$.}. According to (\ref{FS2}) this value saturates the FS prescription for spatially flat FRW models, but it seems fairly incompatible with observational results. However, for spatially closed FRW cosmological models, it has been found that the saturation of the Holographic Principle is related to the value $\omega =-2/3$ which is compatible with current observations (according to \cite{decelini}, $\omega < -0.76$ at the 95\% confidence level). It is likely that the simplest bouncing model (\ref{rclos}) does not describe our universe correctly; however, as shown in this paper, the initial behavior of the universe can enforce the evolution of the particle horizon (future light cone from the beginning) to a saturated scenario compatible with the observed cosmological acceleration\footnote{Work in progress.}. Thus, the dark energy computation based on the Holographic Principle \cite{HDE1,HDE2} seems much more plausible
\begin{equation}\label{DE}
  \rho_{DE} \sim s \, T \sim \frac{S_{PH}/V_{PH}}{D_{PH}} \sim
  \frac{A_{PH}/V_{PH}}{D_{PH}} \sim D_{PH}^{-2} \, .
\end{equation}
Taking $D_{PH} \sim 10 \, \textrm{Gy}$ gives $\rho_{DE} \sim 10^{-10} \,
\textrm{eV}^{4}$ in agreement the measured value \cite{Uflat}.

Finally, two recent conjectures concerning holography in spatially closed universes deserve some comments. W. Zimdahl and D. Pavon \cite{0606555} claim that dynamics of the holographic dark energy in a spatially closed universe could solve the coincidence problem; however the cosmological scale necessary for the definition of the holographic dark energy seems to be incompatible with the particle horizon \cite{HDE1,HDE2,Setare}. In a more recent paper F. Simpson \cite{Simpson} proposed an imaginative mechanism in which the non-monotonic evolution of the particle horizon over a spatially closed universe controls the equation of state of the dark energy. The abundant work in that line is still inconclusive but it seems to be a fairly promising line of work.

\section{Conclusions}

It is usually believed that we live in a very complex and chaotic universe. The Holographic Principle puts a bound for the complexity on our world arguing that a more complex universe would undergo a gravitational collapse. So, one dare say that gravitational interaction is responsible for the simplicity of our world. In this paper a measure of the maximum complexity of the universe compatible with the FS prescription of the Holographic Principle has been deduced. The maximum entropy density at the Planck era under the assumption of a flat FRW universe (\ref{prefac}) and a quintessence dominated spatially closed FRW universe (\ref{prefac2}) has been computed as well.

One of the main points of this paper is to get over an extended prejudice which states that the FS holographic prescription is, in general, incompatible with spatially closed cosmological models. Only two very particular solutions --\cite{KR} and \cite{CruzLepe}-- solved the problem but no physical arguments were given. It has been shown along this paper that cosmological acceleration actually allows the verification of the FS prescription for a wide range of spatially closed cosmological models.

Finally, let us take a further step, a step to a more clear suggestion. First let us assume that the FS prescription is a correct method for the application of the Holographic Principle in Cosmology, then if our universe is spatially closed (although almost flat) it should be accelerated by virtue of the FS prescription. In this sense, the observed acceleration \cite{decelini} enforces the previous assumption. In fact, the experimental results are \emph{compatible with} $k=0$ \cite{Uflat}, but a very small positive curvature cannot be discarded \cite{Uflat5,decelini,Tegm1,0605481}. This \emph{reductionist} use of the Holographic Principle is not usual in the literature. The most common way is to search a valid prescription for every cosmological model and every scenario (like the Bousso solution \cite{Bousso2,Bousso1b}). However, the only possible world we have evidence of is the one which is observed, and maybe it is so because the Holographic Principle does not permit a different one.

\subsubsection*{Acknowledgements}
We acknowledge R. Bousso criticism and suggestions.
This work has been supported by MCYT (Spain) under grant FPA 2003-02948.



\end{document}